\documentclass[11pt, a4paper,sort&compress]{article} 

\usepackage{jcappub}
\usepackage{natbib}
\usepackage{cancel}
\usepackage{pifont} 
\usepackage{amsthm}         
\usepackage{amsmath} 	   
\usepackage{amssymb}       
\usepackage{amsfonts}
\usepackage{graphicx} 	    
\usepackage{verbatim}
\usepackage{color}
\usepackage{colortbl}
\usepackage{float}
\usepackage{multirow}
\usepackage{adjustbox}
\usepackage{mathtools}

\title{Optimal parameterizations for observational constraints on\\thawing dark energy}

\author[a,1]{David Shlivko,%
\note{Corresponding author}}
\author[a]{Paul J. Steinhardt}
\author[b]{and Charles L. Steinhardt}

\affiliation[a]{Department of Physics, Princeton University, Princeton, NJ 08544, USA}
\affiliation[b]{Department of Physics and Astronomy, University of Missouri, 701 S. College Ave., Columbia, MO 65203, USA}

\emailAdd{dshlivko@princeton.edu}
\emailAdd{steinh@princeton.edu}
\emailAdd{csteinhardt@missouri.edu}

\abstract{
Time-varying dark energy is often modeled in observational analyses through generic parameterizations of its equation of state $w(z)$, which typically use two free parameters $\{w_0, w_a\}$ to span a broad range of behaviors as a function of redshift. However, this broad range of behaviors can only approximately capture the dynamics of any given microphysical theory of dark energy.
A complementary approach is to use targeted parameterizations designed to model specific classes of dynamical dark energy with greater precision.
Focusing on the class of thawing dark energy, we quantify and compare the precision with which nineteen generic and targeted parameterizations can capture the dynamics of physically motivated thawing quintessence theories. We find that a targeted parameterization derived from a Pad\'e expansion of $w$ is the most reliable of these, producing accurate reconstructions of $w(z)$, the expansion history $H(z)$, and cosmological parameters such as $H_0$ and $\Omega_m$ for a broad range of microphysical theories. 

}

\arxivnumber{}

\begin{document}
\notoc \maketitle
\flushbottom

\section{Introduction}\label{s_intro}
The possibility that the accelerated expansion of the universe \cite{riess_observational_1998, perlmutter_measurements_1999} is not due to a cosmological constant but, rather, due to some form of dynamical dark energy has motivated a broad range of theoretical explanations. These include canonical ``quintessence'' scalar field theories \cite{Ratra:1987rm,Peebles:1987ek,Wetterich:1987fm,frieman_pngb_1995, Coble:1996te,Turner:1997npq,Caldwell:1997ii}, non-canonical ``k-essence'' theories with tracking behavior \cite{kessence2000, kessence2001}, theories of dark energy rooted in holography \cite{li_model_2004}, and modified gravity (see, e.g., \cite{akrami_modified_2021} for a review). In some theories, the accelerated expansion is eternal; in others, the acceleration ends and transitions to a period of decelerated expansion, which itself either lasts forever or undergoes a second transition to a contracting phase (as in cyclic cosmologies \cite{Ijjas:2019pyf,Ijjas:2021zwv}).

Each of these theories generally predicts a different behavior for the dark energy equation of state $w(z)$, the expansion rate $H(z)$, and other cosmological properties. Combining observations of the cosmic microwave background (CMB), baryon acoustic oscillations (BAO), and Type Ia supernovae (SNe Ia) \cite{Planck:2018vyg, act_extended_2025, DESI:2024vi, desi_cosmo_2025, SN:pantheon, SN:union, SN:DES}, it is possible in principle to determine how well any given theory fits the data and, in particular, whether it fits better than a cosmological constant (as was done for certain quintessence theories in, e.g., refs. \cite{bozek_exploring_2008, akrami_landscape_2019, raveri_swampland_2019, alestas_curve_2024, bhattacharya_cosmological_2024, bhattacharya_cosmological_2025, ramadan_desi_2024, schoneberg_news_2023, dutta_confronting_2007, gupta_observational_2011, ruchika_observational_2023, wolf_scant_2024}).
However, it is impractical to scan the entire space of dark energy theories using such narrow, theory-specific analyses. 
Instead, it is useful to constrain phenomenological parameterizations of $w(z)$ (or, equivalently, $H(z)$) that use just one or two free parameters to distinguish dynamical dark energy theories from each other and from a cosmological constant. 

In choosing such a parameterization, there is a continuous tradeoff between the breadth of theories that can be modeled and the accuracy in modeling any one of them. On one extreme are overly specific parameterizations, which successfully model the dynamics of a small subset of theories but are unable to reproduce the dynamics of others. On the other extreme are broad, generic parameterizations that can very roughly capture the dynamics of many different types of theories. The most common approach in observational analyses today is to use the generic Chevallier-Polarski-Linder (CPL) parameterization \cite{chevallier_accelerating_2001,Linder:2002et,dePutter:2008wt}, which approximates $w$ as evolving linearly with the scale factor $a$ as $w(a) = w_0 + w_a(1-a)$. In terms of redshift, this parameterization translates to
\begin{equation}\label{e_cpl}
w(z)= w_0 + w_a z /(1+z).
\end{equation}
Here, $w_0$ characterizes the equation of state of dark energy at present ($z=0$), and $w_a$ characterizes its rate of change over time. While CPL has the advantage of spanning a broad range of qualitative behaviors ($w(z)$ can be either monotonically increasing or decreasing, and it is not restricted in range), this functional form does not accurately describe the dynamics of any known microphysical theory of dark energy (except for the trivial case of a cosmological constant). 
Due to this inaccuracy, CPL and similar generic parameterizations: 
\begin{enumerate}
\item[(i)] are limited in their ability to distinguish theories of dynamical dark energy from each other and from a cosmological constant \cite{wolf_underdetermination_2023}, 
\item[(ii)] bias inference of cosmological parameters such as $H_0$ and $\Omega_m$, and 
\item[(iii)] may send qualitatively misleading signals about the microphysical nature of dark energy \cite{assessing}.
\end{enumerate}

One way to improve upon the accuracy of CPL is to use a parameterization with more than two degrees of freedom (as in, e.g., refs. \cite{OE_tarrant, Tsujikawa:2013fta, ye_multi, pang_multi, reboucas_multi, escamilla_multi, payeur_multi, rezaei_multi, ribeiro_multi}). However, adding new degrees of freedom comes at the cost of larger error bars on existing parameters, potentially obscured correlations among parameters, and lower statistical significance for a given improvement in fit to observational data \cite{linder_how_2005}.
An alternative way to construct accurate phenomenological parameterizations is simply to restrict the breadth of dark energy dynamics being modeled. Fortunately, there exists a middle ground between narrow, theory-specific analyses and broad but imprecise parameterizations like CPL. 

Within the theory space of dynamical dark energy, there are a handful of classes defined by shared features and behaviors, for which one can construct high-precision parameterizations on a class-by-class basis.
The present work focuses on parameterizing the class of thawing dark energy theories \cite{caldwell_limits_2005, linder_paths_2006, cahn_field_2008, clemson_observational_2009, wolf_scant_2024, payeur_multi, felegary_observational_2024, EXP_camilleri, PADE_alho, akthar_general_2025, adak_generalizing_2014}, in which the equation of state of dark energy is frozen at $w \approx -1$ at early times and increases to larger values during the dark-energy-dominated era. This is arguably the simplest class of dynamical theories of dark energy, as thawing behavior is generic across a broad range of canonical scalar field theories. Our primary goal is to compare various thawing-specific parameterizations (which have a built-in constraint that $w \to -1$ at early times) to each other and to more generic, CPL-like parameterizations, based on how precisely they can capture realistic thawing dark energy dynamics. Our methods for assessing the precision of a parameterization are described in section \ref{s_methods}. The nineteen parameterizations we consider, each using no more than two free parameters, are presented in section \ref{s_params}, and the results of our comparison are described in section \ref{s_results}. In section \ref{s_obs}, we illustrate the use of the best-performing parameterization by placing observational constraints on the phenomenological parameter space and assessing the viability of two sample microphysical theories of thawing dark energy. We conclude with a summary of our results in section \ref{s_discussion}.

Note that throughout this work, we will make reference to two different quantities describing the equation of state of dark energy, namely $w \equiv P/\rho$ (where $P$ is the pressure and $\rho$ is the energy density) and
\begin{equation}
	\epsilon \equiv \frac{3}{2}(1+w).
\end{equation}

\section{Methods for evaluating parameterizations}\label{s_methods}
In this section, we present our approach for evaluating and comparing various parameterizations of thawing dark energy. Given the three drawbacks of CPL outlined in section \ref{s_intro}, an ideal parameterization should be able to simultaneously provide accurate fits to $H(z)$, $H_0$, $\Omega_m$, and $w(z)$ across a broad range of thawing dynamics. To test this broad range of dynamics, each parameterization is fitted to two different physically motivated theories of thawing quintessence (described below in eqs. \ref{e_vexp}-\ref{e_vhill}). 

The fitting process itself optimizes for matching the parameterized expansion history $H_P(z)$ to the theoretical $H_Q(z)$ from a quintessence theory. This approach picks out the best-fit parameters that would be preferred by observational analyses, as $H(z)$ is directly related to key cosmological observables such as Hubble distances, angular-diameter distances, and luminosity distances \cite{assessing}. 
To match the full range of $H(z)$'s probed by observations, a parameterization must be able to precisely capture the shape of the $H(z)$ curve at the low redshifts probed by BAO and SNe Ia ($z \lesssim 4$), as well as the value of the present-day matter density $\rho_m = 3H_0^2\Omega_m$, which fully determines the expansion history $H(z \gtrsim 4)$ during matter domination. In principle, to find the ``best-fit'' parameterization, one has to balance the errors in fitting $\rho_m$ and $H(z \lesssim 4)$ (see, e.g., \cite{wolf_scant_2024, wolf_matching_2025, wolf_robustness_2025}), but this balance depends on the uncertainties in a particular choice of observational data. In this work, we choose to remain more general.

Our approach for comparing parameterizations is to compute the minimum root-mean-square error $E$ between each parameterized expansion history $H_P(z<4)$ and a known thawing-quintessence-driven expansion history $H_Q(z<4)$, giving either zero weight or infinite weight to matching the correct value of $\rho_m$. In the first case, the value of $\rho_m$ is irrelevant, and the error is simply given by
\begin{equation}\label{e_rmserror}
	E_H^\text{min} = \min_{\{\pi_i\}}\sqrt{\frac{1}{4}\int_0^4\left(\frac{H_P(z|\{\pi_i\})-H_Q(z)}{H_Q(z)}\right)^2 dz},
\end{equation}
where $\{\pi_i\}$ is the set of free parameters in the fit. Note that if $P$ parameterizes the equation of state $w(z)$, then computing $H_P(z)$ requires knowledge of not only the one or two parameters in $P$, but also two additional free parameters $\{H_0^P$, $\Omega_m^P\} \subset \{\pi_i\}$. Following the approach of ref. \cite{assessing}, $H_Q(z)$ is computed under the assumption of a flat Friedmann-Robertson-Walker (FRW) universe with a fiducial matter fraction $\Omega_m^Q = 0.3$, using the initial condition $H_0^Q \equiv H_Q(0) = 1$ in arbitrary units. 

In the second case, where the value of $\rho_m$ is matched perfectly, the error is given by
\begin{equation}\label{e_matcherr}
	E_H^\text{max} = \min_{\{\pi_i|\rho_m^P = \rho_m^Q\}}\sqrt{\frac{1}{4}\int_0^4\left(\frac{H_P(z|\{\pi_i\})-H_Q(z)}{H_Q(z)}\right)^2 dz}.
\end{equation}
To enforce a match to $\rho_m$ in practice, we let $H_0^P$ remain a free parameter and then choose $\Omega_m^P$ to satisfy $(H_0^P)^2\Omega_m^P = (H_0^Q)^2\Omega_m^Q = 0.3$. The values of $E_H^\text{min}$ and $E_H^\text{max}$ serve as lower and upper bounds on the error in fitting $H(z<4)$, which together give a sense of the precision with which a given parameterization can model a theory of thawing quintessence. The most robust parameterizations will be able to accurately pick out $\rho_m^P \approx \rho_m^Q$, and thus $E_H^\text{min}$ and $E_H^\text{max}$ will be not only small, but also very similar. More generally, if the $E_H^\text{max}$ of one parameterization is lower than the $E_H^\text{min}$ of another, then the first parameterization is necessarily providing a better fit to the theory.

In order to assess how well the candidate parameterizations can model a broad range of thawing dynamics, they are each fitted to two thawing quintessence theories that represent different types of thawing dynamics: a slow and steady increase in the equation of state on one hand, and a larger, more sudden increase on the other. 
For the slowly thawing quintessence theory, we use an exponential potential,
\begin{equation}\label{e_vexp}
	V_\text{exp}(\varphi) = V_0e^{\lambda \varphi},
\end{equation}
where $\varphi$ is in units of the reduced Planck mass $M_{pl} = \sqrt{\hbar c / (8\pi G)}$, the steepness is set to $\lambda = 1.575M_{pl}^{-1}$, and the initial field value (in the far past, when the dark energy fraction is $\Omega_\varphi = 10^{-6}$) is set to zero. Scalar fields with exponential potentials are ubiquitous in supergravity, modified gravity, and superstring theories; see for example \cite{Wetterich:1994bg,Binetruy:1998rz,Bedroya:2019snp}. 

For the rapidly thawing quintessence theory, we use a quadratic hilltop potential,
\begin{equation}\label{e_vhill}
	V_\text{hill}(\varphi) = V_0\left(1 - \frac{k^2 \varphi^2}{2M_{pl}^2}\right),
\end{equation}
choosing $k^2 = 1$ and setting the initial field value (in the far past, when the dark energy fraction is $\Omega_\varphi = 10^{-6}$) to $\varphi_i = 0.6975 M_{pl}$. Potentials of this type can be used to approximate the dynamics of axion-like fields near the peak of a periodic potential \cite{dutta_hilltop_2008}.  The case of $k^2 = 1$ in particular corresponds to an axion decay constant of $f_a = M_{pl}/(k\sqrt{2}) \approx 0.7 M_{pl}$, allowing for accelerated expansion to be driven without finely tuning of the initial field value and without severely violating the Trans-Planckian Censorship Conjecture \cite{kaloper_pngb_2006, shlivko_tcc}. 
For both the exponential and hilltop theories, the initial field velocity is set to zero, though it is quickly drawn to the attractor trajectory on which the gradient of the potential is balanced by Hubble friction in the scalar field's equation of motion. Additionally, the amplitude $V_0$ in each case is arbitrary, as it drops out of our calculations of $E_H$.

The specific values of $\lambda$ and $\varphi_i$ used in defining these benchmark theories were chosen such that they deviate from the dynamics of a cosmological constant at the level of $E_H^\text{max} \sim 3\%$. 
At this level of deviation, the behavior of thawing dark energy would be measurably different from a cosmological constant, but not so different that it would be definitively ruled out by previous generations of observational data. The errors computed for each parameterization are, of course, going to be specific to this level of deviation, but the ranking of performances should not be too sensitive to this choice.

Once a parameterization is fitted to both benchmark theories through minimization of either $E_H^\text{min}$ or $E_H^\text{max}$, we can compute the corresponding error in capturing the dark energy equation of state at late times for each theory,
\begin{equation}\label{e_ew}
	E_w \equiv \sqrt{\frac{1}{4}\int_0^4\left(w_P(z)-w_Q(z)\right)^2 dz},
\end{equation}
as well as the fractional errors in capturing $H_0$ and $\Omega_m$ for each theory,
\begin{equation}\label{e_h0}
	E_{H_0} \equiv \frac{|H_0^P - H_0^Q|}{H_0^Q}
\end{equation}
and
\begin{equation}\label{e_om}
	E_{\Omega_m} \equiv \frac{|\Omega_m^P - \Omega_m^Q|}{\Omega_m^Q}.
\end{equation}
The candidate parameterizations will ultimately be ranked with an eye toward whether their best-fit errors in $H(z)$, $w(z)$, $H_0$, and $\Omega_m$ are low for both benchmark quintessence theories.

\section{Parameterizations}\label{s_params}
In this section, we introduce the nineteen phenomenological parameterizations of dynamical dark energy that will be tested and compared using the criteria from section \ref{s_methods}. We restrict the scope of this work to parameterizations with no more than two degrees of freedom (dof), and we group these parameterizations into three categories: generic 2-dof parameterizations, thawing 1-dof parameterizations, and thawing 2-dof parameterizations. Some of these parameterizations will be drawn from existing literature, others will be simplified versions of existing parameterizations, and yet others will be new and presented here for the first time. Readers interested in only the top-performing parameterization may skip to section \ref{s_padew}.

\subsection{Generic 2-dof parameterizations}
We define ``generic'' parameterizations of dark energy as those that do not impose, \emph{a priori}, a fixed value for the equation of state $w(z)$ in the early-time limit $z \gg 1$. These parameterizations are designed with the flexibility to capture a broad range of dark energy theories (e.g., thawing, freezing, or phantom dark energy). In this work, we consider five examples of generic 2-dof parameterizations. Three of these parameterize the dark energy equation of state $w(z)$, and their free parameters $\{w_0, w_a\}$ have the same correspondence to $w(0) = w_0$ and $w(\infty) = w_0 + w_a$. The two remaining parameterizations instead model $H(z)$ directly using a generic functional form. Parameterizing $H(z)$ in this way allows for a simple and direct connection between the phenomenological parameters and observational data, but such parameterizations do not generally admit a cosmological-constant limit, and we will show in section \ref{s_results} that they are inferior in precision to the more physically motivated parameterizations of $w(z)$.

\subsubsection{CPL}
The Chevallier-Polarski-Linder (CPL) model \cite{Linder:2002et,chevallier_accelerating_2001, dePutter:2008wt} is the standard 2-dof parameterization for dynamical dark energy used in observational analyses (e.g., refs. \cite{Planck:2018vyg, act_extended_2025, DESI:2024vi, desi_cosmo_2025, SN:pantheon, SN:union, SN:DES}). The equation of state is given by
\begin{equation}
	\boxed{w(z)= w_0 + w_a \frac{z}{1+z}.}
\end{equation}

\subsubsection{BA}
An alternative parameterization from Barboza and Alcaniz \cite{BA_barboza} has been argued to have stronger constraining power than CPL \cite{BA_colgain, BA_giare}. The equation of state is 
\begin{equation}
	\boxed{w(z) = w_0 + w_a \frac{z(1+z)}{1+z^2}.}
\end{equation}

\subsubsection{SQRT}
The square-root parameterization introduced by Pantazis, Nesseris, and Perivolaropoulos has been argued to have a greater quality of fit than CPL and BA to thawing quintessence theories in particular \cite{sqrt_pantazis}. The equation of state is
\begin{equation}
	\boxed{w(z) = w_0 + w_a \frac{z}{\sqrt{1+z^2}}.}
\end{equation}

\subsubsection{CG}
The second-order cosmographic expansion,
\begin{equation}\label{e_cg}
	\boxed{H(z) = H_0(1 + H_1z + H_2z^2),}
\end{equation}
is a model-independent approach to parameterizing the cosmic expansion history. Unfortunately, this series expansion only converges for $|z| \leq 1$, and even then, an accurate approximation generally requires expanding to fourth or fifth order in $z$ \cite{CG_camilleri}. There is also no limit in this parameterization that accurately models the dynamics of a cosmological constant; the closest it can come to matching the $H(z)$'s in $\Lambda$CDM is $E_H^\text{min} = 1.0\%$.

\subsubsection{PADE-L}
It was shown in ref. \cite{Padel_capozziello} that the radius of convergence for parametric fits to $H(z)$ can be extended by replacing the series expansion (\ref{e_cg}) with a (2, 1) Pad\'e approximation of the luminosity distance, from which one can derive
\begin{equation}
	\boxed{H(z) = \frac{(1+z)^2(1+b_1 z)^2}{H_0^{-1} - b_1 H_0^{-1}z^2 + a_2z(2+z+b_1z)}.}
\end{equation}

As with the cosmographic expansion above, this parameterization cannot perfectly model the dynamics of a cosmological constant, though it can come somewhat closer to matching the $H(z)$'s in $\Lambda$CDM, with $E_H^\text{min} = 0.67\%$.

\subsection{Thawing 1-dof parameterizations}
A ``thawing'' parameterization of dark energy is one in which its equation of state is constrained to approach $w(z) = -1$ in the early-time limit $z \gg 1$. Given this constraint, a well-designed thawing 1-dof parameterization should have comparable precision to a generic 2-dof parameterization when fitting a thawing quintessence theory. In this work, we consider five thawing 1-dof parameterizations.

\subsubsection{CPL1}
Imposing $w_0 + w_a = -1$ in the CPL parameterization gives
\begin{equation}
	\boxed{w(z)= w_0 - (1+w_0) \frac{z}{1+z}.}
\end{equation}
The parameter $w_0$ now simultaneously sets the present-day equation of state and the rate of decay toward $w = -1$ as $z \to \infty$. 

\subsubsection{BA1}
Imposing $w_0 + w_a = -1$ in the BA parameterization gives
\begin{equation}
	\boxed{w(z) = w_0 - (1+w_0) \frac{z(1+z)}{1+z^2}.}
\end{equation}

\subsubsection{SQRT1}
Imposing $w_0 + w_a = -1$ in the SQRT parameterization gives
\begin{equation}
	\boxed{w(z) = w_0 - (1+w_0) \frac{z}{\sqrt{1+z^2}}.}
\end{equation}

\subsubsection{EXP1 (simplified)}
Dark energy constraints from refs. \cite{EXP_camilleri, shajib_evolving_2025} used an exponential equation of state,
\begin{equation}\label{e_exp1}
	\boxed{w(z) = -1 + (1+w_0)e^{-1.45 z},}
\end{equation}
to model thawing dynamics. Here, $w_0$ is a free parameter, and fixed value of $1.45$ in the exponent is a slight simplification of the original parameterization, in which this value was allowed to vary within the range $1.45 \pm 0.1$.

\subsubsection{LIN1 (new)}
During matter domination, the equation of state for thawing quintessence obeys \cite{caldwell_limits_2005, cahn_field_2008}
\begin{equation}\label{e_wlin}
	\frac{dw}{dN} = 3(1+w),
\end{equation}
where $N \equiv -\log(1+z)$ counts the number of e-foldings of the scale factor. This behavior is simpler to express using the variable
\begin{equation}
	\epsilon \equiv \frac{3}{2}(1+w), 
\end{equation}
in terms of which eq. (\ref{e_wlin}) becomes
\begin{equation}\label{e_slope3}
	\frac{d\log\epsilon}{dN} = 3.
\end{equation}
In other words, $\log\epsilon$ evolves linearly in $N$ during matter domination with a fixed slope. Extrapolating this behavior into the dark-energy-dominated regime leads to a simple parameterization of thawing quintessence,
\begin{equation}
	\boxed{\log\epsilon(N) = \log \epsilon_0 + 3N,}
\end{equation}
where $\epsilon_0$ is the sole free parameter controlling the equation of state today (at $N = 0$).

\subsection{Thawing 2-dof parameterizations}
Parameterizations with two degrees of freedom on top of the built-in constraint that $w(z = \infty) = -1$ will naturally have greater precision when fitting thawing dark energy theories than either generic 2-dof parameterizations or thawing 1-dof parameterizations. One of our goals will be to quantify this improvement in precision. In this work, we will consider nine thawing 2-dof parameterizations.

\subsubsection{EXP}
This parameterization generalizes EXP1 by allowing variation of the exponent $\alpha$ (within the domain $\alpha > 0$) in the dark energy equation of state
\begin{equation}
	\boxed{w(z) = -1 + (1+w_0)e^{-\alpha z}.}
\end{equation}

\subsubsection{LIN (new)}
This parameterization generalizes LIN1 by allowing variation of the slope $\beta$ in the dark energy equation of state
\begin{equation}\label{e_lin}
	\boxed{\log\epsilon(N) = \log\epsilon_0 + \beta N,}
\end{equation}

where $N \equiv -\log(1+z)$.
Much like the CPL parameterization, LIN is a simple linear fit to the equation of state, except that LIN is defined in log-log space, so that the equation of state never crosses into the ``phantom'' regime ($\epsilon < 0$).

\subsubsection{PWR$\rho$ (new)}
An alternative to the usual approach of parameterizing the dark energy equation of state directly is to parameterize the \emph{energy density} of a thawing quintessence field instead, which begins at some initial asymptotic value and decays at late times. One option is to have the logarithm of the energy density decay as a power law,
\begin{equation}\label{e_pwrrho}
	\log\rho_\text{DE}(N) = \log\rho_0 + \frac{A}{N_{cr}} + \frac{A}{N-N_{cr}},
\end{equation}
where $N \equiv -\log(1+z)$ and $\{A > 0, N_{cr} > 0\}$ are free parameters.
The Friedmann Equations allow this parameterized dark energy density to be translated into the equation of state of dark energy,
\begin{equation}
	\boxed{\epsilon(N) = \frac{A}{2(N-N_{cr})^2}.}
\end{equation}

Curiously, if one were to model the decay of the dark energy density (eq. \ref{e_pwrrho}) with a two-parameter exponential function instead of a power law, one would reproduce the LIN parameterization (eq. \ref{e_lin}).

\subsubsection{AT}
The Algebraic Thawing parameterization from refs. \cite{linder_last_stand, linder_quintessence_dynamics} is given by
\begin{equation}
	\boxed{w(a) = -1 + (1+w_0)a^p\left(\frac{1+b}{1+ba^{-3}}\right)^{1-p/3}.}
\end{equation}
We follow the approach taken in ref. \cite{linder_quintessence_dynamics}, as well as the DESI DR2 analysis of extended dark energy models \cite{desi_extended_2025}, to assume a constant $b = 0.5$ and leave just two free parameters $\{w_0, p\}$. While the DESI analysis restricted their domain to $p > 0$, we allow $p$ in this work to take on any real value. We find that the hilltop and exponential quintessence theories (eqs. \ref{e_vexp}-\ref{e_vhill}) are respectively best-fitted by parameterizations with positive and negative values of $p$.

\subsubsection{HILL}
The hilltop quintessence scenario has been solved analytically in the limit $\epsilon \ll 1$ (equivalently, $w \approx -1$) by Dutta and Scherrer \cite{dutta_hilltop_2008}. Their derived dark energy equation of state,
\begin{equation}
	\boxed{w(z) = -1 + \frac{1+w_0}{(1+z)^{3(K-1)}}\tilde{F}(z)^2,}
\end{equation}
with
\begin{align}
	\tilde{F}(z) &\equiv \frac{(K-F(z))(1+F(z))^K + (K+F(z))(F(z)-1)^K}{(K-\Omega_{\varphi 0}^{-1/2})(1+\Omega_{\varphi 0}^{-1/2})^K+(K+\Omega_{\varphi 0}^{-1/2})(\Omega_{\varphi 0}^{-1/2}-1)^K}, \\
	F(z) &\equiv \sqrt{1+(\Omega_{\varphi 0}^{-1}-1)(1+z)^3},
\end{align}
can in principle be used to model thawing quintessence theories more broadly. The two degrees of freedom in this parameterization are $w_0$ and $K$, while $\Omega_{\varphi 0}$ is set to equal $1 - \Omega_m^P$ in the fitting protocol. Note that this parameterization was used to place observational constraints on thawing dark energy in ref. \cite{chiba_observational_2013}.

\subsubsection{OE (simplified)}
The thawing dark energy equation of state can be parameterized as an ``omega expansion'' (i.e., a series expansion in its fractional energy density) as follows:
\begin{equation}
	\boxed{w(z) = -1 + w_1\Omega_\varphi(z) + w_2\Omega_\varphi(z)^2.}
\end{equation}
Here we have specialized to the case of thawing dark energy by setting the first term to $-1$, rather than treating it as a third degree of freedom as in ref. \cite{OE_tarrant}. Note that unlike the other parameterizations, OE does not give an explicit analytic equation for $w(z)$; instead, one must use the parameters $w_1$ and $w_2$ to first solve for $\Omega_\varphi(z)$. OE does share a similarity with the LIN1 parameterization, though, in that it guarantees the correct early-time evolution of $d\log\epsilon/dN = 3$ during matter domination.

\subsubsection{MUK}
Originally proposed in the context of inflation, Mukhanov \cite{mukhanov_muk} approximated the equation of state of a scalar field driving accelerated expansion as
\begin{equation}
	\boxed{w(N) = -1 + \frac{1 + w_0}{(1-N)^\alpha},}
\end{equation}
where $N \equiv -\ln(1+z)$ and $\alpha > 0$ is enforced to model thawing dynamics. This parameterization was later used in the context of dark energy constraints in ref. \cite{boas_muk}. The two free parameters $w_0$ and $\alpha$ have the usual interpretation of the present-day equation of state and the steepness of decay to $w = -1$ in the far past.

\subsubsection{AH (simplified)}
Akthar and Hossain \cite{akthar_general_2025} have shown that the energy density of a thawing quintessence field can be approximated at late times by
\begin{equation}\label{e_ah}
	\rho_\varphi(a) = \frac{3\Omega_{01}H_0^2}{1 + a^\alpha\left(\frac{\Omega_{01}}{\Omega_{\varphi 0}}-1\right)},
\end{equation}
where we have suppressed a term appearing in the original paper that corresponds to early-time contributions from kinetic energy. Here, $\Omega_{\varphi 0}$ is the present-day dark energy fraction and $\{\alpha > 0, \Omega_{01} > \Omega_{\varphi 0}\}$ are free parameters.
From eq. (\ref{e_ah}), we can use the relation $w(a) = -1 - (1/3) \cdot d\ln\rho_\varphi/d\ln a$ to derive the corresponding equation of state:
\begin{equation}
	\boxed{w(a) = -1 + \frac{\alpha(\Omega_{01}-\Omega_{\varphi 0})a^\alpha}{3(\Omega_{\varphi 0} + (\Omega_{01}-\Omega_{\varphi 0})a^\alpha)}.}
\end{equation}

\subsubsection{PADE-w (simplified)}\label{s_padew}
It was recently demonstrated in ref. \cite{PADE_alho} that the Pad\'e approximant,
\begin{equation}\label{e_pade}
	w(z) = -1 + \frac{(\gamma - 1)\Omega_{\varphi 0}}{\beta \Omega_{\varphi 0}+(1-\gamma \Omega_{\varphi 0})(1+z)^3},
\end{equation}
can model certain thawing quintessence theories to within 0.1\% error, even when the parameters $\Omega_m^P$ and $H_0^P$ are set to their fiducial values rather than being varied to optimize the fit. 
Note that despite being a function of three variables, eq. (\ref{e_pade}) only has two independent degrees of freedom. 
Using the expressions for the present-day equation of state
\begin{equation}
	\epsilon_0 \equiv \frac{3}{2}\left(1+w(0)\right) = \frac{3}{2}\cdot \frac{\gamma - 1}{\beta - \gamma + \Omega_{\varphi 0}^{-1}}
\end{equation}
and the present-day derivative of its logarithm
\begin{equation}
	\eta_0 \equiv \frac{d\ln \epsilon(N)}{dN}|_{N=0} = \frac{3 - 3\gamma\Omega_{\varphi 0}}{1 + (\beta - \gamma)\Omega_{\varphi 0}},
\end{equation}
one can write the dark energy equation of state in the simpler form
\begin{equation}\label{e_pade2}
	\boxed{\epsilon(z) = \frac{3\epsilon_0}{3 + \eta_0 (z^3+3z^2+3z)}.}
\end{equation}
This reparameterization erases the equation of state's dependence on $\Omega_{\varphi 0}$, removes divergences from relevant regions of the parameter space, and uses phenomenological parameters with meaningful physical interpretations.

\section{Comparison of parameterizations}\label{s_results}

\begin{figure*}
	\includegraphics[width=\textwidth]{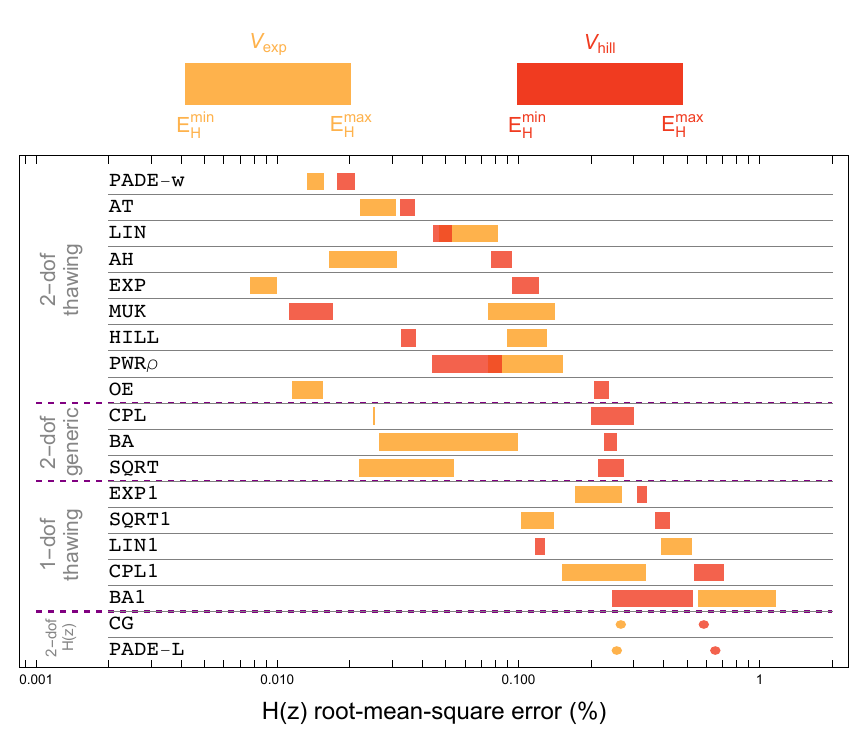}
	\caption{Comparison of errors in fitting $H(z <4)$, as predicted by two different thawing dark energy theories, using nineteen different parameterizations. Rectangles indicate the span of possible errors of the best fit, bounded by $E_H^\text{min}$ and $E_H^\text{max}$ as described in section \ref{s_methods}. Circles are used in cases where $E_H^\text{min} = E_H^\text{max}$. Each parameterization is fitted to the $H(z)$'s from an exponential potential (eq. \ref{e_vexp}, orange) and a quadratic hilltop potential (eq. \ref{e_vhill}, red); the ranked order of parameterizations within each category is determined primarily by comparing the worse of these two fits. While some comparisons are close calls, PADE-w (eq. \ref{e_pade2}) unambiguously has the lowest worst-case and the lowest average errors of all parameterizations considered, with  $E_H^\text{max} \lesssim 0.02\%$ for both benchmark thawing quintessence theories.}\label{f_errors}
\end{figure*}

\begin{figure*}
	\includegraphics[width=\textwidth]{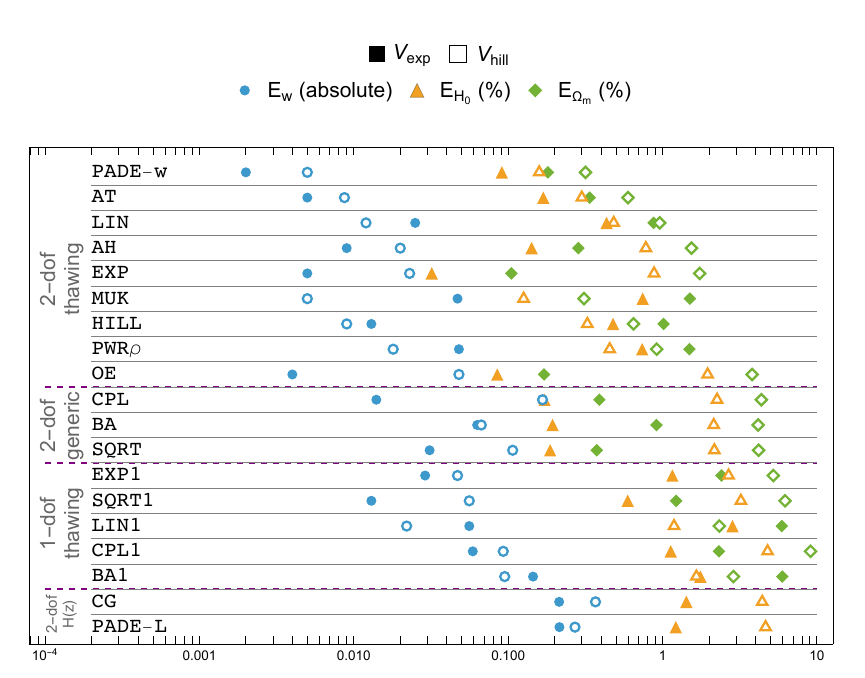}
	\caption{Comparison of absolute errors in $w(z<4)$ (blue circles) and percent errors in $H_0$ (orange triangles) and $\Omega_m$ (green diamonds) using different parameterizations. Each parameterization has been fitted to the $H(z)$'s from an exponential potential (eq. \ref{e_vexp}; solid markers) and a quadratic hilltop potential (eq. \ref{e_vhill}; open markers) to determine the best-fit parameters. The PADE-w parameterization (eq. \ref{e_pade2}) again stands out for having consistently low errors across both quintessence theories fitted, recovering $w(z<4)$ to within $\pm 0.005$ on average and recovering each of $H_0$ and $\Omega_m$ to within $\pm 0.3\%$.}\label{f_errors2}
\end{figure*}

To compare the parameterizations presented in section \ref{s_params}, they were fitted to each of the two benchmark thawing quintessence theories defined by eqs. (\ref{e_vexp}-\ref{e_vhill}). Each of these fittings was performed in two ways, first by optimizing for $E_H^\text{min}$, and second by optimizing for $E_H^\text{max}$, as defined in eqs. (\ref{e_rmserror}-\ref{e_matcherr}). Together, these two optimizations provide a range of possible best-fit errors in fitting the late-time expansion history $H(z<4)$, with larger errors corresponding to tighter early-time constraints on the combination $\Omega_mH_0^2$. We remind the reader that low errors in $H(z)$ are key to fitting cosmological observables and therefore to distinguishing between competing theories of dark energy.

The errors in fitting $H(z)$ are compared in figure \ref{f_errors}, from which one can immediately discern some overarching trends. First, almost all of the 2-dof parameterizations designed to exhibit thawing behavior are able to achieve significantly better fits to thawing-quintessence $H(z)$'s than any of the generic 2-dof parameterizations, especially in the case of the hilltop theory. Second, generic alternatives to CPL with different functional forms --- namely the BA and SQRT parameterizations --- do not generally perform any better than CPL in fitting $H(z)$'s (and in fact perform somewhat worse for the exponential theory). Third, the thawing 1-dof parameterizations performed measurably worse than any 2-dof parameterization of the dark energy equation of state, indicating that at least two free parameters are necessary to precisely capture the breadth of possible thawing dynamics. Finally, our results confirm that parameterizations of $w(z)$ generally produce superior fits to $H(z)$ than direct but overly simplistic parameterizations of $H(z)$ itself (such as CG and PADE-L).

Many of these trends are echoed in figure \ref{f_errors2}, which takes the same fits from above (minimizing $E_H^\text{min}$ or $E_H^\text{max}$) and shows their errors in inferring the dark energy equation of state $w(z<4)$, the Hubble constant $H_0$, and the matter fraction $\Omega_m$. The precise definitions of $E_w$, $E_{H_0}$, and $E_{\Omega_m}$ are given in eqs. (\ref{e_ew}-\ref{e_om}). Because these errors can vary unpredictably depending on whether the fit was optimizing for $E_H^\text{min}$ or $E_H^\text{max}$, we choose to display the larger of the two errors for each point plotted in figure \ref{f_errors2}. One notable difference between figure \ref{f_errors} and figure \ref{f_errors2} is that thawing 1-dof parameterizations can capture thawing $w(z)$'s about as well as generic 2-dof parameterizations, despite having performed worse in fitting $H(z)$'s. This is a reflection of the fact that thawing parameterizations are designed to asymptote to $w = -1$ in the far past; they are therefore guaranteed to have low $w(z)$ errors at large $z$ when fitting thawing quintessence theories. At low $z$, however, there is a broader range of possible thawing dynamics that 1-dof parameterizations cannot fully capture.

Zooming in on the performance of thawing 2-dof parameterizations --- the top-performing category --- we see that PADE-w (eq. \ref{e_pade2}) is the most successful overall based on its consistently low errors, coming in at $E_H^\text{max} \lesssim 0.02\%$ and $E_w \lesssim 0.005$ for both the exponential and hilltop benchmark theories.
The other parameterizations in this category either have consistent but larger errors when fitting the exponential and hilltop theories (like AT, LIN and PWR$\rho$), or they perform noticeably worse for one thawing potential than the other. EXP and OE, for example, outperform even PADE-w when fitting the exponential-potential $H(z)$'s, but they do almost as poorly as CPL when modeling hilltop dynamics. The reverse is true for MUK. We note that the performance of the AT parameterization, which nearly matches that of PADE-w, relies on allowing negative values for the parameter $p$ (which were not considered for the recent DESI analysis \cite{desi_extended_2025}).

External testing indicates that PADE-w continues to perform exceptionally at modeling the thawing dynamics of potentials with different shapes as well, such as steeper quadratic hilltops, quartic hilltops, and the plateau potentials from refs. \cite{Andrei:2022rhi, assessing}. Moreover, when modeling dynamics of scenarios with larger $\Omega_m^Q > 0.3$ (as one can expect to infer in thawing quintessence scenarios due to the dilution of dark energy at late times), the errors of the PADE-w parameterization are even lower, as long as $w_Q(z=0) < 0$. This makes our ranking of parameterizations robust to changes in $\Omega_m^Q$ within this moderate thawing regime.

On the other hand, when modeling more extreme thawing scenarios with $w_Q(z=0) > 0$, a given value of $\Omega_m^Q$ will be reached twice: once during the $w_Q < 0$ regime during which $\Omega_m$ is decreasing, and then again during the $w_Q > 0$ regime as $\Omega_m$ increases. In these scenarios, larger values of $\Omega_m^Q$ correspond to larger values of $w_Q(z=0)$ and will generally yield larger best-fit errors for any parameterization. As a result, generic parameterizations become even more severely inaccurate and the advantages of targeted parameterizations are amplified; however, the relative rankings of targeted 2-dof parameterizations in this regime may vary depending on the specific dynamics of the quintessence theory. The errors of PADE-w in particular reach $E_H^\text{max} \approx 0.09\%$ for the benchmark hilltop theory when reaching $\Omega_m^Q = 0.36$ with $w_Q(z=0) \approx 2.3$ and $E_H^\text{max} \approx 0.46\%$ for a steeper hilltop with $k^2 = 20$ reaching $\Omega_m^Q = 0.36$ with $w_Q(z=0) \approx 4.2$. (For reference, the corresponding CPL errors in these cases are in excess of $1\%$.)

\section{Observational constraints}\label{s_obs}
In this section, we use the PADE-w parameterization to derive analogous constraints to those found in the most recent DESI analysis using the AT parameterization \cite{desi_extended_2025}. We also demonstrate how these phenomenological constraints can be propagated into constraints on microphysical theories of thawing dark energy. Our analysis will use a combination of three datasets:
\begin{enumerate}
	\item \textbf{CMB.} For low-$\ell$ temperature (TT) and polarization (EE) power spectra, we use respectively the Commander and SimAll likelihoods from the 2018 Planck PR3 release \cite{planck_2020_like}. For high-$\ell$ temperature (TT), polarization (EE) and cross-spectra (TE), we use the Python-native lite (nuisance-marginalized) Plik likelihood from the same release. 
	We also use a combination of CMB lensing data from the Atacama Cosmology Telescope \cite{ACT_lensing1, ACT_lensing2, ACT_lensing3} and Planck's PR4 (NPIPE) maps \cite{carron_Planck_lensing}, choosing the \texttt{actplanck\_baseline} likelihood variant \cite{act_lensing_git}.
	\item \textbf{BAO.} We use the DESI DR2 observations of galaxies, quasars, and the Lyman-alpha forest, measuring the angular diameter distance $D_M(z)$, the Hubble distance $D_H(z)$, or the angle-averaged quantity $D_V(z) \equiv (zD_M(z)^2D_H(z))^{1/3}$ in seven redshift bins. The individual DESI observations span $0.1 \leq z \leq 4.2$, while the binned effective redshifts range from $z = 0.295$ to $z = 2.330$ \cite{desi_forest_2025, desi_cosmo_2025}. 
	\item \textbf{SNe Ia.} 
		We use the Union3 compilation of 2087 SNe Ia from redshifts $0.01 < z < 2.26$ \cite{SN:union}. The twenty-two binned effective redshifts range from $z = 0.05$ to $z = 2.26$.
\end{enumerate}

We perform our analyses with \texttt{cobaya} \cite{cobaya1, cobaya2}, making use of the \texttt{CAMB} Boltzmann solver \cite{camb1, camb2} and the Metropolis-Hastings MCMC sampler with dragging \cite{metropolis1, metropolis2, metropolis_drag}. Within \texttt{CAMB}, we use the parameterized post-Friedmann approach \cite{fang_ppf} to compute dark energy perturbations, and we modify the 2016 \texttt{HMcode} algorithm \cite{mead_hmcode_2015, mead_accurate_2016} for computing non-linear matter power spectra to be compatible with a non-CPL equation-of-state parameterization. 
As is the default in \texttt{cobaya}, the CMB lensing anomaly $A_\text{planck}$ is treated as a nuisance parameter with a gaussian prior ($\mu=1$, $\sigma=0.0025$). The convergence of MCMC chains is determined using the Gelman-Rubin statistic \cite{gelman_inference_1992}, for which we require $R - 1 < 0.01$. Finally, we analyze the MCMC results using \texttt{GetDist} \cite{getdist}, discarding the first 20\% of the chains as burn-in. Our analyses assume zero spatial curvature and a single massive neutrino, with $m_\nu = 0.06$ eV and $N_\text{eff} = 3.044$ total neutrino species. 

Variable cosmological parameters and their priors are listed in Table \ref{t_priors}. For ease of comparison, we assume uniform priors and upper bounds on $\epsilon_0$ and $\eta_0$ that represent a similar range of dynamics to the AT parameter space used in ref. \cite{desi_extended_2025}. However, we note that these priors place a disproportionate amount of weight on regions of the parameter space where the equation of state is ultra-rapidly varying at late times, corresponding to $\eta_0 \gg 1$. (For reference, the exponential and hilltop theories used as fitting benchmarks respectively have $\eta_0 \approx 0.7$ and $\eta_0 \approx 4.7$, though the PADE-w parameterization can still accurately model steeper hilltops corresponding to $\eta_0 \sim \mathcal{O}(100)$, with $E_H^\text{max} \lesssim 0.06\%$.) We will also see that the upper bounds on $\epsilon_0$ and $\eta_0$ are saturated by the posterior distribution, so one should interpret the results of these analyses with due caution.

\begin{table*}
\begin{center}
	\begin{tabular}{|l|l|}
		\hline
		Parameter & Prior\\
		\hline
		$\ln(10^{10}A_s)$ & $\mathcal{U}[1.61, 3.91]$ \\
		$n_s$ & $\mathcal{U}[0.8, 1.2]$\\
		$\tau$ & $\mathcal{U}[0.01, 0.8]$\\
		$100\theta_\text{MC}$ & $\mathcal{U}[0.5, 10]$\\
		$\omega_b$ & $\mathcal{U}[0.005, 0.1]$\\
		$\omega_c$ & $\mathcal{U}[0.001, 0.99]$\\
		$\epsilon_0$ & $\mathcal{U}[0, 3]$\\
		$\eta_0$ & $\mathcal{U}[0, 100]$\\
		\hline
	\end{tabular}
	\caption{Parameters and priors used in the MCMC analyses. Here, $A_s$ is the amplitude of the primordial power spectrum, $n_s$ is the spectral index, $\tau$ is the optical depth, and $\theta_\text{MC}$ is the angular size of the sound horizon at recombination. The physical densities of baryons (b) and cold dark matter (c) are defined as $\omega_i \equiv \Omega_i h^2$, with $h \equiv H_0 / (100\text{ km/s/Mpc})$. The dark energy equation of state is defined by the PADE-w parameters $\{\epsilon_0, \eta_0\}$ per eq. (\ref{e_pade2}).}\label{t_priors}
\end{center}
\end{table*}

\begin{figure*}
	\includegraphics[width=0.529\textwidth]{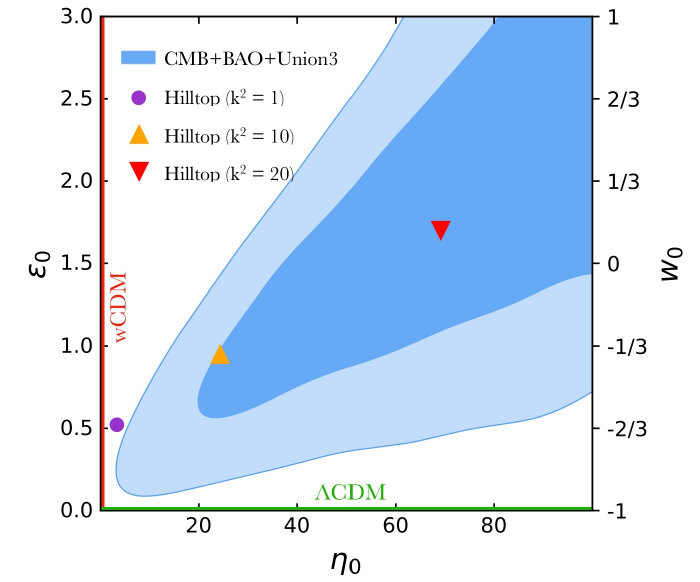}
	\includegraphics[width=0.46\textwidth]{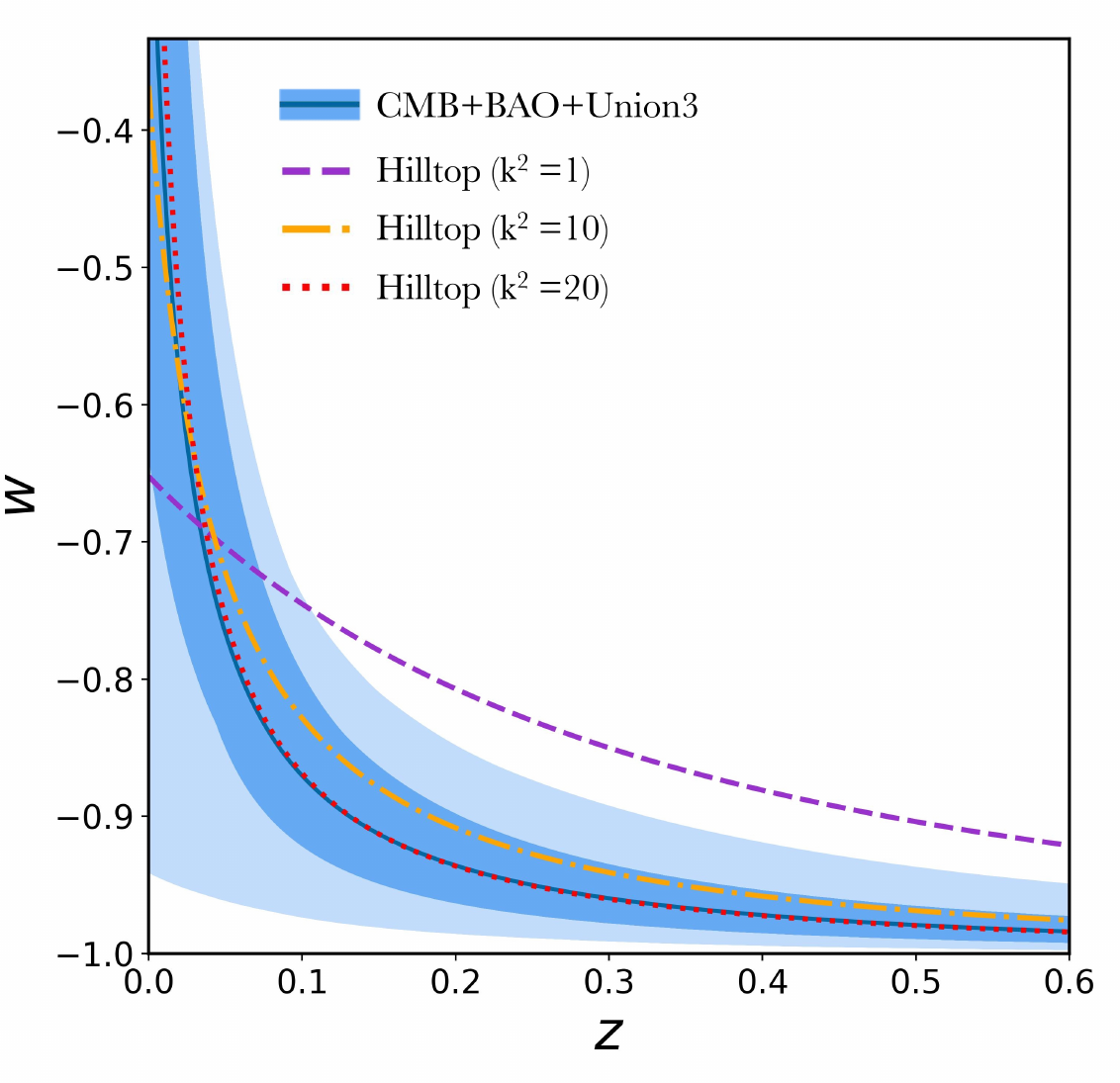}
	\caption{\emph{Left panel}: MCMC posteriors showing $1\sigma$ and $2\sigma$ constraints on the PADE-w parameters $\{\epsilon_0, \eta_0\}$. Limits of the parameter space corresponding to $\Lambda$CDM ($w = -1$) and wCDM ($w = $ const.) are labeled and highlighted. Also shown are the best-fit points representing three sample hilltop quintessence theories with $V(\varphi)/V_0 = 1 - k^2\varphi^2/(2M_{pl}^2)$. A shallow hilltop (purple circle) with $k^2 = 1$ and $\varphi_i = 0.66M_{pl}$ is excluded at $2\sigma$; a steeper hilltop (orange triangle) with $k^2 = 10$ and $\varphi_i =  0.0245M_{pl}$ is just barely compatible with the $1\sigma$ constraint, and the steepest hilltop (red downward-pointing triangle) with $k^2 = 20$ and $\varphi_i = 0.004M_{pl}$ is well within the $1\sigma$ contour. \newline \emph{Right panel}: Best-fit evolution of $w(z)$ with $1\sigma$ and $2\sigma$ confidence regions, as reconstructed from the PADE-w constraints. Overlaid are the $w(z)$ curves predicted by the same three hilltop quintessence theories, assuming $\Omega_m = 0.3$. As in the left panel, the shallower hilltop (purple, dashed) is ruled out at $2\sigma$, while the steeper hilltop (orange, dot-dashed) lies just barely within the $1\sigma$ contour and the steepest hilltop (red, dotted) very nearly matches the best-fit evolution of $w(z)$. In both panels, one sees that the data allow for the possibility of present-day deceleration, with $w_0 = w(z=0) > -1/3$.}\label{f_flatpriors}
\end{figure*}

The joint constraints on $\epsilon_0$ and $\eta_0$ are shown in the left panel of figure \ref{f_flatpriors}. The posteriors are in tension with $\Lambda$CDM at the level of roughly $2.2\sigma$, with $\Delta \chi^2_\text{MAP} \approx -7$ and $\Delta$DIC $\approx -4.2$ (in line with the findings of ref. \cite{desi_extended_2025}). 
The posteriors are overlaid with best-fit points corresponding to three different hilltop quintessence theories, as determined using the approach of ref. \cite{assessing} to minimize $E_H^\text{min}$. There is a noticeable preference for steeper hilltop potentials, given the current data and choice of prior; this is consistent with previous results from, e.g., ref. \cite{wolf_scant_2024}. The right panel of figure \ref{f_flatpriors} shows the evolution of $w(z)$ reconstructed from the $1\sigma$ and $2\sigma$ confidence contours in the PADE-w parameter space, and this reconstruction is overlaid with predictions of $w(z)$ from the same three hilltop quintessence theories. Comparing these overlaid predictions to the reconstructed evolution, one can draw the same conclusion that the steeper hilltops are preferred over the shallower ones, with consistent quantitative levels of statistical significance.

The ability to constrain microphysical theories of dark energy by directly comparing predicted to reconstructed $w(z)$ curves is a key advantage of high-precision targeted parameterizations. Any parameterization that cannot accurately model the $w(z)$'s of a particular theory would give misleading results when assessing the viability of that theory through a reconstruction of $w(z)$. This is why constraints using the CPL parameterization, for example, can only be translated into constraints on microphysical theories of dark energy indirectly through matching $H(z)$'s \cite{assessing}.

\section{Conclusions}\label{s_discussion}

In this work, we have evaluated and compared the precision with which nineteen parameterizations of dark energy can model the dynamics of thawing quintessence theories. These include both generic parameterizations and targeted parameterizations with thawing dynamics ($w \to -1$ at early times) built in. Our key result is that the (targeted) PADE-w parameterization,
\begin{equation}
	\epsilon(z) \equiv \frac{3}{2}(1+w(z)) = \frac{3\epsilon_0}{3 + \eta_0 (z^3+3z^2+3z)},
\end{equation}
stands out for its remarkable precision in capturing the expansion history $H(z)$, the dark energy equation of state $w(z)$, and the cosmological parameters $H_0$ and $\Omega_m$ across a broad range of theoretical thawing-dark-energy scenarios. The performance of PADE-w is visualized in comparison to other parameterizations in figures \ref{f_errors}-\ref{f_errors2}.

In section \ref{s_intro}, we outlined three drawbacks of generic dark energy parameterizations such as CPL. By using a targeted 2-dof parameterization like PADE-w instead, all three drawbacks are ameliorated for analyses of thawing dark energy:
\begin{enumerate}
\item[(i)] Precise fits to $H(z)$ (and therefore to key cosmological observables) improve the statistical significance with which theories of thawing dark energy can be distinguished from each other and from a cosmological constant;
\item[(ii)] Precise fits to the shape of $H(z)$ also allow for more accurate inference of cosmological parameters such as $H_0$ and $\Omega_m$ in thawing-dark-energy scenarios; and 
\item[(iii)] Reliable inference of $w(z)$ makes it straightforward to translate constraints on the phenomenological parameter space into constraints on microphysical dark energy theories.
\end{enumerate}

To elaborate on point (iii), we showed in section \ref{s_obs} that the reconstruction of $w(z)$ from constraints on the PADE-w parameters can be used to constrain theories of dark energy by direct comparison to their predicted $w(z)$ curves. In contrast, generic parameterizations like CPL can only be used to constrain theories of dark energy indirectly, as their reconstructions of $w(z)$ do not represent any realistic microphysical dynamics \cite{assessing}.



Of course, the drawback of using targeted parameterizations lies in their limited scope. Fits to $H(z)$ and inference of $w(z)$, $H_0$, and $\Omega_m$ using the PADE-w parameterization may be highly inaccurate, for example, if the true dynamics of dark energy in the universe are not well-described by a thawing equation of state. It is therefore important to assess statistically whether the PADE-w thawing parameterization provides a better or worse fit to the data compared to generic parameterizations like CPL and compared to targeted parameterizations optimized for other physically motivated classes of dark energy. We look forward to presenting these comparisons in future work.

 \vspace{0.1in}
\noindent
{\it Acknowledgements.} 
We thank A. Mead, C. Meyerhoff, G. Montefalcone, P. Phillips, E. Witten, and R. Wojtak for useful conversations. 
The MCMC simulations used in this paper were performed on computational resources managed and supported by Princeton Research Computing, a consortium of groups including the Princeton Institute for Computational Science and Engineering (PICSciE) and the Office of Information Technology's High Performance Computing Center and Visualization Laboratory at Princeton University.
 DS and PJS are supported in part by the DOE grant number DEFG02-91ER40671 and by the Simons Foundation grant number 654561. 

\bibliographystyle{JHEP.bst}
\bibliography{thawing.bib}

\end{document}